\documentclass{article}

\usepackage{amsmath}
\usepackage{amssymb}
\usepackage{graphicx}
\usepackage{bbm}

\def\R{{\mathbbm{R}}}
\def\S{{\mathbb{S}}}
\def\d{{\mathrm{d}}}

\title{Topological approach to phase transitions and inequivalence of statistical ensembles}

\author{Michael Kastner\footnote{e-mail: {\tt Michael.Kastner@uni-bayreuth.de}}\\ \small Physikalisches Institut, Lehrstuhl f\"ur Theoretische Physik I,\\ \small Universit\"at Bayreuth, 95440 Bayreuth, Germany}

\date{}

\begin{document}


\maketitle

\begin{abstract}
The relation between thermodynamic phase transitions in classical systems and topology changes in their state space is discussed for systems in which equivalence of statistical ensembles does not hold. As an example, the spherical model with mean field-type interactions is considered. Exact results for microcanonical and canonical quantities are compared with topological properties of a certain family of submanifolds of the state space. Due to the observed ensemble inequivalence, a close relation is expected to exist only between the topological approach and {\em one}\/ of the statistical ensembles. It is found that the observed topology changes can be interpreted meaningfully when compared to microcanonical quantities. 
\end{abstract}

Phase transitions, like the boiling and evaporating of water at a certain temperature and pressure, are common phenomena both in everyday life and in almost any branch of physics. Loosely speaking, a phase transition brings about a sudden change of the macroscopic properties of a system while smoothly varying a parameter (the temperature or the pressure in the above example). For the description of equilibrium phase transitions within the framework of statistical mechanics, several so-called statistical ensembles or Gibbs ensembles, like the microcanonical or the canonical one, are at disposal, each corresponding to a different physical situation. For a large class of systems with sufficiently short ranged interactions, these different approaches lead to identical numerical values for the typical system observables of interest, after taking the thermodynamic limit of the number of particles in the system going to infinity \cite{Ruelle}. In this situation one speaks of equivalence of ensembles. Then, instead of selecting the statistical ensemble according to the physical situation of interest, one can revert to the ensemble most convenient for the computation intended. For systems with long range interactions, however, equivalence of ensembles does not hold in general. Systems showing such an inequivalence of ensembles in the thermodynamic limit (among those gravitational systems and Bose-Einstein condensates) have attracted much research interest in the last years (see Ref.\ \cite{Dauxois_etal} for a review). Dramatic differences between the ensembles can be observed for example in the specific heat, which is a strictly positive quantity in the canonical ensemble, whereas negative values, and even negative divergences, can occur in the microcanonical ensemble \cite{Eddington,LynWood}.

An entirely different approach to phase transitions, not making use of any of the Gibbs ensembles, has been proposed recently. This {\em topological approach}\/ connects the occurrence of a phase transition to certain properties of the potential energy $V$, resorting to {\em topological}\/ concepts. From a conceptual point of view, this approach has a remarkable property: The microscopic Hamiltonian dynamics can be linked via the Lyapunov exponents to the topological quantities considered \cite{CaPeCo1}. With the topological approach, in turn, linking a change of the topology to the occurrence of a phase transition, a concept is established which provides a connection between a phase transition in a system and its underlying microscopic dynamics.

The topological approach is based on the hypothesis \cite{CaCaClePe} that phase transitions are related to topology changes of submanifolds $\Sigma_v$ of the state space of the system, where the $\Sigma_v$ consist of all points $q$ of the state space for which $V(q)/N=v$, i.\,e., their potential energy per degree of freedom equals a certain level $v$. (Or, in a related version, the topology of submanifolds $M_v$ consisting of all points $q$ with $V(q)/N\leqslant v$ is considered.) This hypothesis has been corroborated by numerical and by exact analytical results for a model showing a first-order phase transition \cite{Angelani_etal1,Angelani_etal2} as well as for systems with second-order phase transitions \cite{CaPeCo1,CaCoPe,CaPeCo2,GriMo,RiTeiSta,Kastner}. A major achievement in the field is the recent proof of a theorem, stating, loosely speaking, that, for systems described by smooth, finite-range, and confining potentials, a topology change of the submanifolds $\Sigma_v$ is a {\em necessary}\/ criterion for a phase transition to take place \cite{FraPeSpi}.

Albeit necessary, such a topology change is clearly not {\em sufficient}\/ to entail a phase transition. This follows for example from the analytical computation of topological invariants in the XY model \cite{CaCoPe,CaPeCo2}, where the number of topology changes occurring is shown to be of order $N$, but only a single phase transition takes place. So topology changes appear to be rather common, and only particular ones are related to phase transitions. There are strong indications that a criterion based exclusively on topological quantities cannot exist in general \cite{Kastner}.

Having mentioned the recent efforts to more firmly establish sufficient and necessary relations between topology changes and phase transitions, and bearing in mind the phenomenon of ensemble inequivalence, we notice an additional complication: In the case of inequivalence, such a relation cannot simply connect topology changes to phase transitions, but, because of the existing differences, only to phase transitions {\em in a certain statistical ensemble}. It is the objective of the present Letter to clarify the connection between the topological approach and the description of phase transitions within the various statistical ensembles.

To this purpose, the so-called spherical model with mean field-type interactions is considered, a model in which the inequivalence of ensembles is of a peculiar kind. After briefly introducing this model, its thermodynamic behavior is discussed, making use of the microcanonical as well as of the canonical approach. A temperature driven phase transition is found to occur in the canonical framework, whereas no transition is present microcanonically.\footnote{The terms ``microcanonical'' and ``canonical'' have been used in connection with the spherical model in a different context in the literature \cite{YanWan}, referring to the question whether the constraint (\ref{constraint}) holds strictly (``microcanonically''), or only on average (``canonically''). This is not to be confused with the traditional meaning of those terms that we refer to.} Confronting these results with the topology of the state space submanifolds $\Sigma_v$ of this model, the close connection between the topological approach and the microcanonical ensemble becomes evident. From these findings, conclusions about the topological approach by A.\ C.\ Ribeiro Teixeira and D.\ A.\ Stariolo \cite{RiTeiSta} are discussed. Finally, the close connection between the topological approach and the microcanonical ensemble is addressed on a more general level.

{\em Mean field spherical model}.--- Introduced by T.\ H.\ Berlin and M.\ Kac \cite{BerKac} in 1952, the spherical model of a ferromagnet is devised such as to mimic some features of the Ising model, while, at the same time, having a continuous state space and being exactly solvable in the thermodynamic limit for arbitrary spatial dimension of the lattice. We consider a mean field-like simplification of the original model where, instead of nearest-neighbor interactions, all degrees of freedom interact with each other at equal strength. The unitless Hamiltonian function of this model is given by\footnote{A slightly different definition in which the first sum in the Hamiltonian extends over all $i,j$ with $i\neq j$ is used in \cite{RiTeiSta}. This merely results in a shift of the energy scale: $v\to v+\frac{1}{2N}$.}

\begin{equation}
H(\sigma)=-\frac{1}{2N}\sum_{i,j=1}^N \sigma_i \sigma_j - h\sum_{i=1}^N \sigma_i,
\end{equation}
where $h$ is an external magnetic field and the $N$ degrees of freedom $\sigma_i\in\R$ ($i=1,...,N$) are subject to the additional constraint
\begin{equation}\label{constraint}
\sum_{i=1}^N \sigma_i^2=N.
\end{equation}
This constraint restricts the space of allowed states $\sigma=(\sigma_1,...,\sigma_N)$ to an ($N$--1)-sphere $\S^{N-1}\subset\R^N$ with radius $\sqrt{N}$.

{\em Microcanonical ensemble}.--- The starting point to describe the spherical model in the microcanonical ensemble is the density of states
\begin{equation}\label{Omega}
\Omega_N(v)=A_N^{-1}\idotsint\limits_{\sum\limits_{i=1}^N \sigma_i^2=N} \d\sigma_1...\d\sigma_N \,\delta[H(\sigma)-Nv],
\end{equation}
where $\delta$ denotes the Dirac distribution and $A_N$ is a normalization constant \cite{BerKac}. By a change of variables, the integral in (\ref{Omega}) can be deformed into the integral over the surface of an $(N-2)$-sphere (or two such spheres, depending on the values of $v$ and $h$ considered). Performing the thermodynamic limit, the microcanonical entropy
\begin{align}
s(v)&=\lim_{N\to\infty}\frac{1}{N}\ln \Omega_N(v)\nonumber\\
&=\frac{1}{2}\ln\left[1-\left(|h|-\sqrt{h^2-2v}\right)^2\right],\label{s_v}
\end{align}
is obtained. This function is defined for all $v$ which, for a given external field $h$, ensure a positive argument of the square root and of the logarithm. For arbitrary fixed $h$, the entropy $s$ is a smooth function on its entire domain, and therefore no phase transition occurs in the microcanonical ensemble.

Despite its simplicity, to the best of our knowledge, expression (\ref{s_v}) has not yet been reported in the literature. But even more can be done: the spherical mean-field model is one of the rare models with interacting degrees of freedom, for which the microcanonical entropy can be computed explicitely for systems with an arbitrary finite number $N$ of degrees of freedom \cite{Kastner_inprep}.

{\em Canonical ensemble}.--- G.\ S.\ Joyce \cite{Joyce} has given an exact canonical solution for the spherical model with quite general long range interactions in the thermodynamic limit. For the mean field-like case of infinite range interactions, the internal energy $v$ per degree of freedom as a function of the inverse temperature $\beta=\frac{1}{T}\in\R_0^+$ (with Boltzmann's constant $k_B\equiv1$) can also be obtained by a Legendre-Fenchel transform from (\ref{s_v}). For the case of zero external field $h=0$, we obtain
\begin{equation}
v(\beta)=\begin{cases}
0 & \text{for $\beta\leqslant1$,}\\
\frac{1-\beta}{2\beta} & \text{for $\beta>1$.}
\end{cases}
\end{equation}
For non-zero external field $h$, the internal energy $v$ is given by the real root of a third order polynomial in $v$,
\begin{equation}
8\beta^2 v^3-4\beta(\beta h^2-2\beta+2)v^2-2(6\beta^2 h^2-\beta^2-2\beta h^2+2\beta-1)v+\beta h^2(4\beta h^2-\beta+2)=0.
\end{equation}
Note that, for $h\neq0$, the internal energy $v$ is a smooth function of $\beta$ on $\R_0^+$, whereas, for zero field $h$, a cusp at $\beta=1$ is present, implying the occurrence of a temperature driven phase transition in the latter case [see \cite{RiTeiSta} for exemplary plots of the graph of $v(\beta)$].

{\em Ensemble inequivalence}.--- For non-zero external field, no phase transition occurs in the mean field spherical model, neither in the microcanonical, nor in the canonical ensemble. In fact, the canonical internal energy $v(\beta)$ turns out to be the inverse function of the microcanonical inverse temperature $\beta(v)=\frac{\partial s(v)}{\partial v}$ for positive $\beta$. For zero external field $h=0$, however, remarkable differences between the two ensembles are observed: The {\em absence}\/ of a temperature (or energy) driven phase transition in the microcanonical ensemble is in contrast to the {\em presence}\/ of such a transition in the canonical ensemble. The particular scenario occurring is termed {\em partial equivalence of ensembles}\/ in \cite{ElHaTur}, and a few other models of statistical mechanics are known to exhibit this phenomenon \cite{Barre_etal}.

\begin{figure}[hbt]
\begin{center}
\includegraphics[width=7.2cm,height=6.6cm,clip=false]{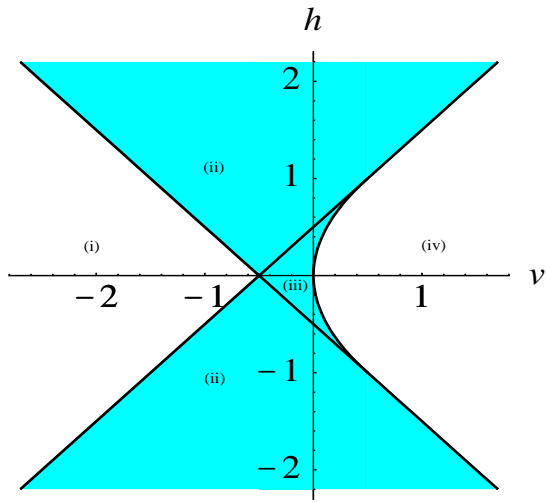}
\caption{\label{changes} \small
Lines in the $(v,h)$-plane where topology changes occur. The shaded region marks the support of the density of states $\Omega_N(v)$, i.e., the values of $v$ which, for a given external field $h$, are accessible for the system. The various regions in the plane separated by the bold lines correspond to the following submanifold topologies: (i) to the left of the shaded region: $\Sigma_v\sim\varnothing$, (ii) the two shaded triangles: $\Sigma_v\sim\S^{N-2}$, (iii) the small, triangle shaped shaded region:  $\Sigma_v\sim\S^{N-2}+\S^{N-2}$, (iv) to the right of the shaded region: $\Sigma_v\sim\varnothing$.}
\end{center}
\end{figure}

{\em Topological approach}.--- As a starting point for the topological approach, the submanifolds $\Sigma_v$ of the state space, consisting of all points where the potential energy per particle equals $v$, are considered. In the case of the spherical model, the Hamiltonian $H$ consists of a potential energy term only, and we define
\begin{equation}\label{Sigma_def}
\Sigma_v=\left\{ q\in\Gamma\,\Big|\,\frac{H(q)}{N}= v\right\},
\end{equation}
where the state space $\Gamma$ of the spherical model is an ($N$--1)-sphere $\S^{N-1}$ with radius $\sqrt{N}$. Ribeiro Teixeira and Stariolo have analyzed the topological structure of the state space submanifolds of the mean field spherical model in \cite{RiTeiSta}. Completing their findings and translating them into our notation, the topology of the submanifolds $\Sigma_v$ can be written as
\begin{equation}\label{Sigma_topo}
\Sigma_v\sim
\begin{cases}
\S^{N-2} & \text{for $-|h|-\frac{1}{2}<v<|h|-\frac{1}{2}$},\\
\S^{N-2}+\S^{N-2} & \text{for $|h|-\frac{1}{2}<v<\frac{h^2}{2}$ and $|h|<1$},\\
\varnothing & \text{else},
\end{cases}
\end{equation}
where $\sim$ indicates topological equivalence, the $+$-sign denotes a topological (unconnected) sum, and $\varnothing$ is the empty set. The regions of the $(v,h)$-plane corresponding to the different topologies and the lines at which topology changes occur are plotted in Figure \ref{changes}.

{\em Discussion of the results}.--- Our intention is to investigate the connection between the topology of the state space submanifolds $\Sigma_v$ and thermodynamic quantities of the system. In light of the fact that microcanonical and canonical results are not equivalent, we can expect a necessary and sufficient relation to exist, if at all, only between the topological approach and {\em one}\/ of the statistical ensembles.

Comparing the topology of $\Sigma_v$ in (\ref{Sigma_topo}) with the {\em canonical}\/ thermodynamics of the mean field spherical model with zero external field $h=0$, one would associate the phase transition observed at inverse temperature $\beta=1$ and internal energy $v=0$ with the topology change from $\S^{N-2}+\S^{N-2}$ to $\varnothing$ at $(v,h)=(0,0)$. However, the same type of topology change is present for all $|h|<1$ at $(v,h)=(\frac{h^2}{2},h)$, whereas canonical thermodynamics asserts the absence of a phase transition for all $h\neq0$. From this observation, and disregarding the presence of ensemble inequivalence in the mean field spherical model, Ribeiro Teixeira and Stariolo \cite{RiTeiSta} concluded that the information contained in the topology of $\Sigma_v$ (or, in a related description, of $M_v$) might not be sufficient as to distinguish between the presence of a phase transition or its absence.

However, a much simpler explanation arises from a comparison of the topology of $\Sigma_v$ with the {\em microcanonical}\/ thermodynamics of the model. Microcanonically, the zero field case is not special, as a phase transition occurs neither with nor without an external field. This is in accordance with the topological results where the topology change is the same for all $|h|<1$.

Note that the mean-field spherical model is not the only example for which the observed topology changes find a meaningful interpretation only when compared to microcanonical quantities. For the mean-field XY model considered in \cite{CaCoPe,CaPeCo2}, the observed discontinuity in the modulus of the Euler characteristic is interpreted as being related to the phase transition in the canonical ensemble at $h=0$. However, similarly to the mean-field spherical model, the same topology change is present for $h\neq0$, although a phase transition does not occur in this case. Again, a comparison with microcanonical quantities can provide a meaningful interpretation of the findings.\footnote{Note, however, that for a class of systems with {\em non-confining}\/ potentials the information from the submanifold topology was found to be insufficient as to distinguish between the occurence and the absence of a phase transition \cite{Kastner}.}

After having discussed the relation between the topological approach and the two ensembles (the microcanonical and the canonical one, respectively) on the basis of examples, some general remarks are in order. The microcanonical ensemble is the most fundamental of the Gibbs statistical ensembles: it is based only on the assumption of equal {\em a priori}\/ probabilities of the states of the system, and the other statistical ensembles can be derived from it. The topological approach, in turn, is a fundamental description in the sense that it is closely connected to the microscopic dynamics which is the basic level of description underlying the thermodynamics of the system. Intuitively it seems reasonable to expect a closer relationship between these two particularly fundamental approaches.

This relation can be substantiated by noting that the microcanonical ensemble and the topological approach, different as they are, have the same quantity as a starting point: The microcanonical partition function (=density of states) of a system, which is at the basis of the microcanonical ensemble, can be defined as the {\em volume}\/ of the level sets (\ref{Sigma_def}). Considering, instead of the volume, the {\em topology}\/ of the very same level sets, we arrive at the fundamental quantity underlying the topological approach, and, in the light of this observation, the close connection between these two concept appears plausible.

{\em Summary}.--- We have presented exact results for microcanonical, canonical, and topological quantities of the mean field spherical model. Comparing the microcanonical and the canonical quantities, we observe that, for this model, equivalence of the statistical ensembles does not hold: microcanonically, no phase transition occurs, whereas in the canonical ensemble a phase transition takes place for zero external field. In order to investigate the connection between the thermodynamics of the system and the topological approach to phase transitions, we have compared topological quantities to microcanonical and canonical ones. As a consequence of the observed ensemble inequivalence, a one-to-one connection can, at best, be expected between the topological approach and {\em one}\/ of the statistical ensembles. From the results for the mean field spherical model, we find that it is the comparison of topological quantities with {\em microcanonical}\/ results which allows a meaningful and convincing interpretation. The close conceptual connection between the topological approach and the microcanonical ensemble suggests that, whenever ensemble inequivalence occurs, microcanonical thermodynamics is the adequate reference frame to confront the topological approach with.

\section*{Acknowledgments}
I would like to thank Oliver Schnetz for his help with the mathematics%
. Financial support by the Deutsche Forschungsgemeinschaft (grant KA2272/2) is gratefully acknowledged.

\end{document}